\documentclass[sigconf,authorversion=true]{acmart}

\usepackage{subcaption}
\usepackage{float}
\usepackage{natbib}

\AtBeginDocument{%
  }

\copyrightyear{2023}
\acmYear{2023}
\setcopyright{acmlicensed}\acmConference[ARES 2023]{The 18th International Conference on Availability, Reliability and Security}{August 29-September 1, 2023}{Benevento, Italy}
\acmBooktitle{The 18th International Conference on Availability, Reliability and Security (ARES 2023), August 29-September 1, 2023, Benevento, Italy}
\acmPrice{15.00}
\acmDOI{10.1145/3600160.3605000}
\acmISBN{979-8-4007-0772-8/23/08}

\begin{document}

%
% The "title" command has an optional parameter,
% allowing the author to define a "short title" to be used in page headers.
\title{Cookiescanner: An Automated Tool for Detecting and Evaluating GDPR Consent Notices on Websites}

%
% The "author" command and its associated commands are used to define
% the authors and their affiliations.
% Of note is the shared affiliation of the first two authors, and the
% "authornote" and "authornotemark" commands
% used to denote shared contribution to the research.
\author{Ralf Gundelach}
\email{rg.psi@uni-bamberg.de}
\orcid{0009-0006-8745-4583}
\affiliation{
  \institution{University of Bamberg}
  \city{Bamberg}
  \country{Germany}
}

\author{Dominik Herrmann}
\email{dh.psi@uni-bamberg.de}
\orcid{0000-0002-7374-3054}
\affiliation{
  \institution{University of Bamberg}
  \city{Bamberg}
  \country{Germany}
}

\begin{abstract}
%Currently 240 words, 200-250 words max.
%Erster Satz weg
%The General Data Protection Regulation (GDPR) has changed the internet landscape, setting out strict guidelines that website providers must meet to collect and process user data legally. 
%In terms of the quality of the extracted result, the BERT language model performs best because it is text-based and detects all additional text and buttons, even if they are visually separated in the banner. 
The enforcement of the GDPR led to the widespread adoption of consent notices, colloquially known as \emph{cookie banners}. 
Studies have shown that many website operators do not comply with the law and track users prior to any interaction with the consent notice, or attempt to trick users into giving consent through dark patterns. 
Previous research has relied on manually curated filter lists or automated detection methods limited to a subset of websites, making research on GDPR compliance of consent notices tedious or limited. 
We present \emph{cookiescanner}, an automated scanning tool that detects and extracts consent notices via various methods and checks if they offer a decline option or use color diversion.  
We evaluated cookiescanner on a random sample of the top 10,000 websites listed by Tranco. 
We found that manually curated filter lists have the highest precision but recall fewer consent notices than our keyword-based methods. 
Our BERT model achieves high precision for English notices, which is in line with previous work, but suffers from low recall due to insufficient candidate extraction. 
While the automated detection of decline options proved to be challenging due to the dynamic nature of many sites, detecting instances of different colors of the buttons was successful in most cases. 
Besides systematically evaluating our various detection techniques, we have manually annotated 1,000 websites to provide a ground-truth baseline, which has not existed previously. Furthermore, we release our code and the annotated dataset in the interest of reproducibility and repeatability. 
\end{abstract}





\maketitle

\section{Introduction}\label{sec:introduction}

%The GDPR has been a milestone in establishing consumer rights by giving them greater control over their data and establishing strict requirements for valid consent companies must fulfill to process consumer data lawfully. 
To comply with the GDPR, website providers are using consent notices, colloquially known as \emph{cookie banners}. These pop-up messages appear when a user first visits a website, informing them of the use of cookies and similar technologies to track their browsing behavior. Ideally, these consent notices provide simple options to accept, decline, or set more granular settings for tracking cookies. 

While there has been an increase in the overall number of notices since the enforcement of the GDPR~\cite{degeling_we_2019}, the presence of a notice alone does not make a website provider GDPR-compliant: 
Besides apparent violations, such as tracking the user before any chance to object~\cite{papadogiannakis_user_2021} or ignoring the user's  decision, many websites feature limited or no options to object, or use notice designs that make objecting as difficult as possible~\cite{matte_cookie_2020, sanchez-rola_can_2019}. 

The latter practices fall under the category of ``Dark Patterns''~\cite{harry_brignull_httpswwwdeceptivedesign_2022}, UI design elements that coerce users towards undesired outcomes. 
Prior works are limited to manual and semi-automatic approaches~\cite{degeling_we_2019, matte_cookie_2020, sanchez-rola_can_2019, soe_circumvention_2020}. There have been some attempts at automating the detection of consent notices but these only cover a subset of consent notices captured through automated filtering lists~\cite{eijk_impact_2019, kampanos_accept_2021, sheil_fianan_2022} or third-party consent management platforms (CMPs)~\cite{bollinger_automating_2022, nouwens_dark_2020, papadogiannakis_user_2021}. 
Furthermore, these detection methods were used without examining error rates, introducing unknown amounts of bias into the results. 

In addition to evaluating the performance of the detection methods, we also attempt to determine whether the consent notice offers an option to decline and whether the notice uses the same color for the options or tries to draw the attention of the user towards certain buttons. 
Moreover, we advance the field by presenting a systematic evaluation approach that was applied to a sample of 1,000 websites. For details, see Section~\ref{sec:results}. 
% We scanned the top 10,000 websites in the Tranco list as of 2023-03-01 and manually analyzed a random sample of 1,000 websites. 

To summarize, we offer the following contributions:
\begin{enumerate}
\item We present \emph{cookiescanner}, an automated scanning tool to detect and interact with consent notices. 
\emph{cookiescanner} allows us to conduct insightful comparisons of various detection methods. 
We evaluate four different detection methods for 1,000 popular websites.  
\item Furthermore, \emph{cookiescanner} allows us to detect simple dark patterns such as ``obstruction'', i.\,e., not providing an option to decline on the first layer of the notice, and ``interface interference'' in the form of different colored buttons. Again, we evaluate our approach for 1,000 websites. 
\item We also publish our detection engine and the datasets analyzed in the paper~\cite{gundelach_cookiescanner_2023}. This includes the machine learning classifier, the dataset used to train it, and the consent notice screenshots from our ground truth. This allows subsequent research to reproduce our results and enables future work. The screenshots could also be used to train an image-based classifier. 
\end{enumerate}

The rest of this work is structured as follows: 
Section~\ref{sec:background} provides insight into the reason behind the widespread adoption of consent notices and the criteria a notice has to fulfill to allow the user to make an informed decision. 
Section~\ref{sec:related_works} discusses related work on the detection of consent notices, dark patterns in consent notices, and the effectiveness of dark patterns in consent notices.  
Section~\ref{sec:methodology} explains the implementation of our scanner and the different detection methods. 
Section~\ref{sec:results} presents the results of our scan. 
Section~\ref{sec:discussion} discusses the results and our limitations. Section~\ref{sec:conclusion} presents a conclusion highlighting opportunities for further research.
 
\section{Background}\label{sec:background}

Directive 2002/58/EC, also known as ePrivacy directive, sets out fundamental rules for processing personal data and protecting privacy in the electronic communications sector within the European Union. 
The directive applies to all forms of electronic communications, including email, telephone, and internet services. It establishes specific rules on the usage of cookies and similar technologies, unsolicited communications (spam), and location data. 
Furthermore, it requires website providers to obtain consent and inform users of the purpose of cookies before setting them on their devices. 
It also states that giving and withholding consent to data collection practices should be as user-friendly as possible~\cite{publications_office_of_the_european_union_directive_2002}. 

Directive 2009/136/EC, also known as the Cookie Directive, extends this by requiring website providers to ask for permission and inform users about the purpose of the cookies set on the device of the user if they serve other purposes, such as tracking and are not necessary for the basic functionality of the website~\cite{publications_office_of_the_european_union_directive_2009}.

The most recent legislative effort is the GDPR~\cite{publications_office_of_the_european_union_regulation_2016}, which came into force in 2018. The GDPR aims to provide a uniform framework for protecting and processing personal data in the European Union. 
In particular, it gives users extensive rights over their data. It prohibits processing personal data, except for specific circumstances, unless the user has explicitly consented and has been well-informed about the purpose.  
The GDPR applies to all website providers, including those outside the EU, if they process the data of EU citizens. 
Notably, it also introduces hefty fines for non-compliance, which has led to the widespread adoption of consent notices. 

The GDPR provides guidelines on what constitutes valid consent. Under the GDPR, consent needs to be freely given (without any coercion or pressure from the data controller), specific (for a particular purpose and not bundled with other requests), informed (clear and understandable information about what the user is consenting to, including who processes the data, how the processor retains the data and whom he shares it with), unambiguous (explicit action, such as ticking a box, clicking a button, or signing a document) and revocable at any time, while giving consent should be as easy as withdrawing consent. 

However, it must be noted that there is no common legal framework and that these decisions are made on a case-by-case basis by the responsible DPA or court, making it impossible to derive a universally applicable consent notice design based on the legislation. 
Some courts and DPAs have taken action against pre-ticked checkboxes~\cite{court_of_justice_of_the_european_union_case_2019} and website providers failing to provide a reject button~\cite{commission_nationale_de_linformatique_et_des_libertes_cookies_2022} on the first layer of the banner. 

In addition, the European Data Protection Board (EDPB) set up a \emph{cookie banner task force}, which published a report highlighting several practices that the EDPB believes make it impossible for a notice to obtain valid consent. 
In terms of notice design, the report cites the lack of a reject button, pre-ticked check boxes, misleading design that presents the option to decline as a link, and designs that use color and contrast to highlight the option to accept~\cite{european_data_protection_board_report_2023}.

\section{Related Works}\label{sec:related_works}

The following paragraphs describe work relevant to our research.
We cover methods for detecting consent notices, detecting dark patterns within consent notices, and experiments investigating the effectiveness of different dark patterns within consent notices. 

\subsection{Detection of Consent Notices}

Van Eijk et al.~\cite{eijk_impact_2019} used a crowd-sourced filter list containing CSS selectors to identify consent notices. 
They report a False Positive Rate (FPR) of 1\% and a False Negative Rate (FNR) of 20\% during a crawl of 1,500 websites. 

Coudert~\cite{coudert_automatically_2020} developed a three-step algorithm: First, he segments the HTML tree, assigns scores using a keyword list to the individual segments, and acquires the consent notice by traversing the HTML tree of the highest scored element downwards until the score decreases. 
However, despite the introduction of a threshold value to limit the number of false positives (FP), Coudert still reports up to a third of the approximately 10,000 detected notices as FPs after scanning the top 100,000 sites on the Tranco list. 

Hausner and Gertz~\cite{hausner_dark_2021} broadly discuss their consent notice detection algorithm which is based on keyword matching. They describe extracting the clickable elements based on the HTML tags and applying text-based clustering to differentiate between the various clickable elements. 
However, they do not present any data on the performance of their approach. 

Khandelwal et al.~\cite{khandelwal_cookieenforcer_2022} offer a more promising approach. They use the stacking order of the elements in the page (e.\,g.,  a positive z-index) to identify candidate elements which they evaluate using a pre-trained BERT classifier. 
They only report one FP in a manually verified set of 500 pages from the Tranco top list. Similarly, Bouhoula uses the presence of the keyword ``cookies'' and a positive z-score for his pattern matching, which yields no FP, but a lower detection ratio compared to Khandelwal et al. ~\cite{bouhoula_automated_2022}. 

Klein et al.~\cite{klein_accept_2022} used a combination of keywords in different languages and visual properties (visible and whether it is at the top layer of the screen) to identify elements that notices use to get consent (e.\,g., accept buttons). 
The main goal of their research was not to detect consent notices but to measure the security impact of loaded third-party code in terms of client-side cross-site scripting. 
In a random sample of 250 pages from the top 10,000 of the Tranco list, their approach correctly identified the consent button, while missing 48 notices.

\subsection{Dark Patterns in Consent Notices}

Gray et al. conducted an expert analysis of a manually collected corpus of websites and categorized dark patterns according to five strategies they pursue: 
Nagging (redirecting of expected functionality), obstruction (making a process more difficult), sneaking (attempting to hide information from the user), interface interference (manipulation of the user interface to highlight or hide certain elements) and finally, forced action (requiring a user to perform a specific action to access certain functionality)~\cite{gray_dark_2018}. Current research and court rulings provide ample evidence for the presence of dark patterns as shown in the following. 
For instance, many websites used to provide pre-ticked checkboxes, which has been found to be unlawful~\cite{publications_office_of_the_european_union_case_2019}. 

Nouwens et al.~\cite{nouwens_dark_2020} scraped the consent notices of the top five consent management platforms in the UK (QuantCast, OneTrust, Cookiebot, and Crownspeak) on the top 10,000 websites according to the Alexa ranking. About 50.1\,\% of the scraped notices featured no ``Reject All'' button, while in 74.3\,\% of the cases, the ``Reject All'' button was hidden one layer deep, i.\,e., it became only visible once users had clicked on a button in the first layer of the consent notice. 
This practice is not in line with the notion of informed consent and recently resulted in a 150 million Euro fine for Google~\cite{commission_nationale_de_linformatique_et_des_libertes_cookies_2022}. 

Soe et al. manually examined cookie consent notices from 300 news outlets. 
The most common dark patterns they found were Obstruction in 43\,\% and Interface Interference in 45.3\,\% of the consent notices~\cite{soe_circumvention_2020}. 

\subsection{The Efficacy of Dark Patterns in Consent Notices}

Given the prevalence of dark patterns on websites, this raises the question of their effectiveness. 
Utz et al. conducted a series of field experiments and determined that defaults in the form of pre-selected checkboxes and highlighted buttons were an effective nudging tool~\cite{utz_informed_2019}. 
In contrast to that, Graßl et al.~\cite{grasl_dark_2021} were not able to confirm the effect of color highlighting. 
They reason that due to their ubiquitous nature and frequent exposure, users have become accustomed to this type of dark pattern. 
Berens et al.~\cite{berens_cookie_2022} confirmed the findings of Graßl et al. regarding color highlighting and additionally discovered that styling the decline option as a link instead of a button had a strong effect. 

\section{Methodology}\label{sec:methodology}

First, we describe four different detection methods to detect consent notices. 
After that, we describe the design and implementation of our scanner. 
Finally, we describe our scanning methodology and how we selected and evaluated our final dataset. 

\subsection{Method 1: DOM Tree Walking}

We observed that almost all consent notices contain the word ``cookie'' at least once. 
The data collection and annotation of the dataset for our machine learning approach confirms this assumption. 
In the top 1,000 websites of the Tranco list as of 2023-01-01, only 1\,\% of all English consent notices did not contain the word ``cookie''.  
Of the non-English consent notices, only one did not contain the keyword ``cookie'', but this notice referred to the privacy legislation of the respective country, not the GDPR. 

Our approach works as follows: 
First, we search the DOM for all nodes containing the keyword ``cookie''. 
Then we extract their node id, the word length of the text, and their x and y coordinates. 
Of these nodes, we remove all that are outside the boundaries of the viewport of the browser and select the one with the longest text. 
Starting from this node, we traverse the DOM upwards until we reach the node below the body tag. 
However, this led to us going beyond our target and capturing additional elements of the website, like the whole header or footer. We optimized our strategy by stopping when we were able to extract a button or when the next parent node is the body tag. 

\subsection{Method 2: Perceptive Detection}

\begin{figure*}
  %\centering
  \captionsetup[subfigure]{aboveskip=-2em}
  \begin{subfigure}{0.49\textwidth}
  	\includegraphics[width=\textwidth]{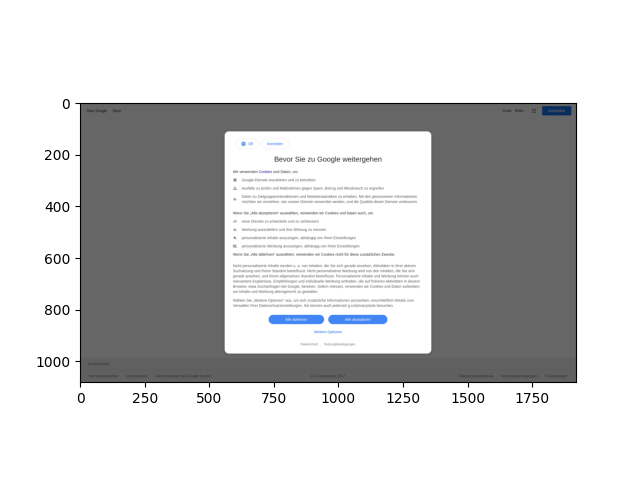}
  	\caption{Page Screenshot}
  	\Description{A screenshot of the website https://google.com}
    \label{fig:page-screenshot}
  \end{subfigure}
  %\hfill
  \begin{subfigure}{0.49\textwidth}
  	\includegraphics[width=\textwidth]{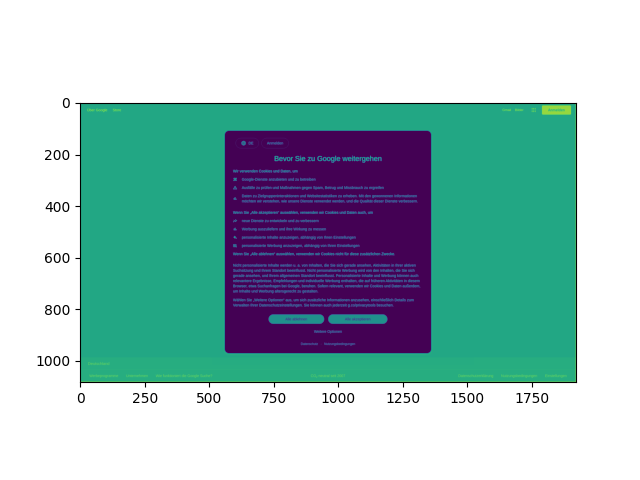}
    \caption{Page Screenshot XORed with Background Color}
    \Description{A screenshot of the website https://google.com where the image has ben XORed with the color behind the keyword cookie, showing a negative of the consent notice}
    \label{fig:page-screenshot-xor}
  \end{subfigure}
  \begin{subfigure}{0.49\textwidth}
  	\includegraphics[width=\textwidth]{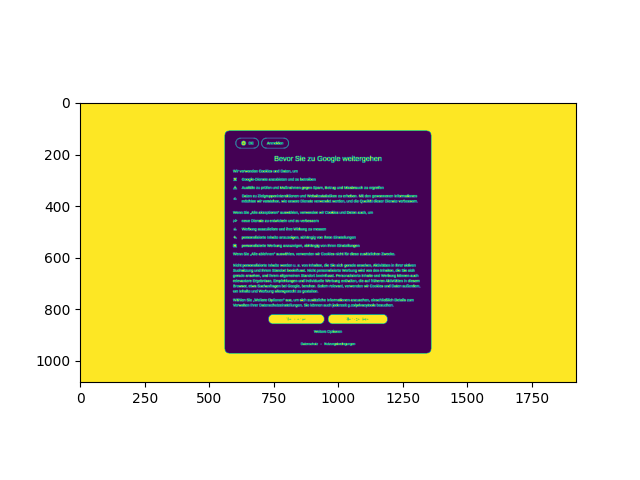}
    \caption{Page Screenshot XORed and Grey Scale}
    \Description{Negative image of the consent notice that has been converted to grey-scale}
    \label{fig:page-screenshot-xor-grey-scale}
  \end{subfigure}
  %\hfill
  \begin{subfigure}{0.49\textwidth}
  	\includegraphics[width=\textwidth]{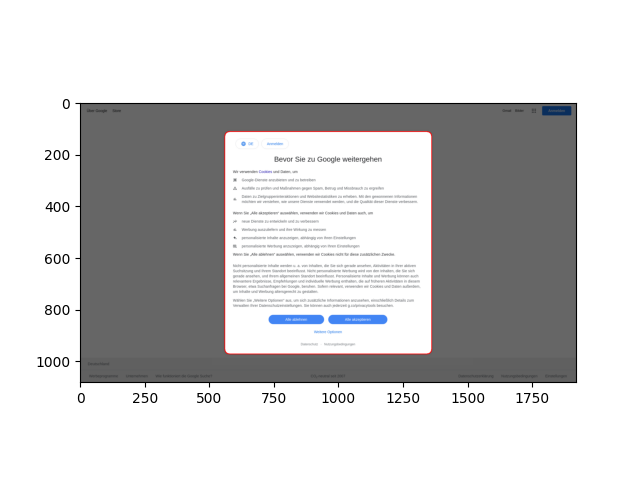}
    \caption{Page Screenshot with Highlighted Consent Notice}
    \Description{Negative image of the consent notice that has been converted to grey-scale}
    \label{fig:page-screenshot-contour-highlight}
  \end{subfigure}
  \caption{Image Processing of the Perceptive Detection}
  \label{fig:perceptive-detection}
\end{figure*}

In contrast to Method 1, where we rely on the heuristic (the presence of a button), Method 2 uses image processing techniques to determine the boundaries of the notice. For that purpose, we consider the background color behind the keyword (e.\,g., ``cookie''). Figure~\ref{fig:perceptive-detection} shows all image processing steps. 

First, the scanner removes all embedded images from the web page, takes a screenshot of the website and places a black border of one pixel around the screenshot (Figure~\ref{fig:page-screenshot}). The border is necessary so that we are able to capture the contour of consent notices spanning the whole width of the page. 
Then, we extract the background color behind the chosen keyword and XOR an image filled with pixels in the background color of the notice with the screenshot of the website to obtain a negative of the consent notice (Figure~\ref{fig:page-screenshot-xor}). 
We convert the result of the XOR operation to grey-scale and obtain a binary (black and white) image via thresholding (Figure~\ref{fig:page-screenshot-xor-grey-scale}).
Then we use OpenCV~\cite{bradski_opencv_2000} to retrieve a list of all the contours and select the smallest one that still contains the coordinates of the keyword. 
Figure~\ref{fig:page-screenshot-contour-highlight} shows the screenshot of the web page with the chosen contour of the consent notice highlighted in red. 
Finally, we retrieve the DOM node at the location of the first pixel within the contour. 

Since the node at this location could be a part of the notice (i.\,e., the body of the notice without buttons and headlines), we further optimize our result:  
We traverse the DOM upwards while the area of the node increases and is less than or equal to that of the extracted contour, and the dimensions of the node are still within the bounds of the contour. The reason for this is so that we do not traverse outside of the consent notice. 

\subsection{Method 3: List-based Detection}

This detection method identifies consent notices by matching the DOM of a webpage against CSS selectors from the manually maintained filter lists ``EasyList''~\cite{easylist_filter_list_project_easylist_2023} and ``I don't care about Cookies''~\cite{kladnik_i_2023}. 
These lists originally serve the purpose of removing consent notices without breaking the website. It would be interesting to know if they are a reliable method for detecting consent notices, as some previous work relied on ``I don't care about Cookies''~\cite{eijk_impact_2019, kampanos_accept_2021, sheil_fianan_2022}. 

\subsection{Method 4: BERT-based Detection}

The BERT Detection collects candidate elements according to the methodology described by Khandelwal et al.~\cite{khandelwal_cookieenforcer_2022}. 
The candidate elements are a list of: 
\begin{enumerate}
\item All visible elements with a positive z-index, i.\,e., those that are higher in the stacking order and less likely to be covered by other elements and
\item The first three and last three text nodes of the HTML body. 
\end{enumerate}
The detector loops through all candidate elements and lets a BERT model ~\cite{devlin_bert_2019} classify whether the text of the candidate elements fits the text of a consent notice.  
The first element that is classified as a consent notice is considered to be the notice. 
\begin{table}[ht]
\caption{\label{tab:bert_classifier} Performance of the BERT classifier}
\begin{tabular}{l c c c c}
\toprule
\textbf{Instances} 	& \textbf{Support}	& \textbf{Recall}	& \textbf{Precision}	& \textbf{F1}\\
\midrule
Notices				& 50 			    & 0.98				& 1.00				& 0.99\\
Candidate Elements	& 50				    & 1.00				& 0.98				& 0.99\\
\midrule
Total				& 100				& 0.98				& 1.00				& 0.99\\
\bottomrule
\end{tabular}
\end{table}

We reached out to Khandelwal et al. and asked whether they could provide their dataset or trained model but received no response.  
Therefore, we trained our own model. To gather training and testing data, we manually annotated the top 1,000 websites of the Tranco~\cite{le_pochat_tranco_2019} list of 2023-01-01. 
Via manual inspection, we were able to extract a total of 1877 candidate elements and 361 notices in the English language. 
Additionally, we gathered 65 consent notices and 212 candidate elements in German. Given the small number of elements in German, we decided to only train an English model. 

We used the following parameters, matching the parameters of Khandelwal et al. as closely as possible. 
We chose a BERT case-based model with a learning rate of $2e^{-5}$, AdamW as the optimization algorithm, and trained for ten epochs with batches of eight, using the last checkpoint for evaluation. 
We used a balanced training set of 300 notices and 300 non-notices, and 50 notices and 50 non-notices for evaluation. 
All entries in the datasets were randomly selected and excluded from the random draws of the other datasets to ensure that there was no overlap between training and evaluation data. 
Table~\ref{tab:bert_classifier} shows the results of our evaluation: 
We achieved a precision of 1.00 and a recall of 0.98, with only one notice misclassified as a non-notice. 
It was the consent notice from facebook.com, one of the longest in the dataset, which contained the term ``cookie'' a total of 23 times. 
Overall, our results are slightly better than the results reported by Khandelwal et al.~\cite{khandelwal_cookieenforcer_2022}, achieving an F1 score of 0.99 compared to their 0.97. 

\subsection{Design and Operation of the Scanner}

This section provides a high-level overview of the implementation of the scanner and the steps it carries out to scan a website. 
The scanner consists of a Python program that spawns multiple workers. 
Each of these workers instruments a Chromium instance via the ChromeDevTools API. 
The scanner retrieves the websites to scan from a PostgreSQL database and stores the results in JSON-format in the aforementioned database. 

\paragraph{Stage 1: Consent Notice Detection}

First, the worker loads the web page and waits five seconds before taking action. 
If the worker does not receive a response or cannot resolve the domain during a 60-second timeout, it logs the error and aborts the scan. 
Throughout the scan, the worker logs all set cookies while recording HTTP requests and responses through callback functions. 
After successfully loading the page, the worker saves a screenshot of the web page and attempts to identify consent notices using the methods described in the sections before. 
For each consent notice found, the worker extracts the properties of the consent notice (relevant for the analysis: position, html, text, width, height, fontsize, language as detected with langdetect~\cite{danilk_langdetect_2021} and takes a screenshot of the consent notice and a screenshot of the web page with the consent notice highlighted. 

\paragraph{Stage 2: Interacting with the Consent Notice}

The input of Stage~2 is a DOM element that contains a notice as identified in Stage~1. Since we have executed all four detection methods in Stage 1, we may have obtained conflicting information about the presence and location of the notice. In the following, we explain how we handle this situation for the purpose of this paper.

For sake of conciseness and reproducibility, we chose to run Stage~2 only on a single detection method's result. If a result was provided by BERT Detection, we use that result. Otherwise we resort to the result returned by Perceptive Detection, DOM Tree Walking, and List-based Detection in this order.

More exhaustive analyses are left for future work. We note, however, that exhaustive analyses that involve reloading pages multiple times will trigger DDoS protection measures on some websites, which have to be detected and handled.

For each clickable element, the worker clears the browser cache and local memory and reloads the page. After waiting, the worker retrieves the newly appeared notice using the detector from the priority list. 
For the re-scan, the worker does not record properties and screenshots of the notice. 
The worker then fetches the clickable element by text (going by node ID alone proved to be unreliable due to possible changes in the DOM after page reload), clicks it, and waits five seconds. 
The worker then saves all requests, responses, and cookies before and after clicking the element and saves a screenshot of the page after the click. 

Finally, the worker uses the screenshot to calculate the structural similarity index measure (SSIM)~\cite{wang_image_2004} of the screenshot after the click and the screenshot during the initial page load. 
The aim of comparing the screenshots was to determine the presence of a decline option by checking whether two screenshots lead to the same result (i.\,e., causing the notice to disappear). We choose this approach because it is agnostic of the text or content of links and buttons. While it would be simpler to rely on the text of the clickable elements (e.\,g., by searching for the string ``reject all''), this would result in a keyword-based detector that is tedious to maintain and not very robust. 

Determining whether an element is clickable and what role it plays proved to be quite difficult, since web design is quite variable and query selectors are unable to capture all possible variations of buttons, checkboxes, and links. 
First, we use JavaScript to determine whether the mouse pointer changes when hovering over an element. 
If this is true, the scanner considers the element to be a clickable. 
Then, we check if the element contains child nodes and iterate over all child nodes. If the node itself or its child nodes contain the ``checked'' attribute or the accessibility attribute ``aria-checked'', we consider it to be a checkbox and extract its value. 
Otherwise, we check whether it contains the ``href'' attribute and whether it leads to a new web page, since we have observed at least one CMP with a higher market share and other websites using ``href'' attributes with paragraphs as buttons, as well as some web pages executing JavaScript in the ``href'' attribute. 
Finally, we consider all clickables that do not match the above criteria buttons. 

\paragraph{Stage 3: Detecting Interface Interference and Obstruction}

%Our scanner attempts to detect two types of dark patterns: Interface interference (manipulation of the user interface to highlight or hide certain elements) and obstruction (making a process more difficult than it should be, e.\,g.,  providing an opt-out option only in the second layer of the banner). 

In the following, we describe how the scanner analyzes the notice. 

The scanner determines the color of a button by counting the pixels of each color of the clickable element and determining the dominant color by majority vote. Initially, we tried extracting the color via JavaScript through the computed style of the element, but this proved to be unreliable. 

If the screenshots obtained after clicking on two different buttons are identical according to the structural similarity metric, this means that both buttons lead to the same state. 
Based on our experience from manual inspection, this is always the case when there is an accept and a decline option in a consent notice, both of which result in closing the banner and displaying the actual website.

%The design of the consent notice determines the presence of obstruction. Ideally, a banner would feature an `Accept'' and ``Decline'' of equal precedence and, optionally, a third button featuring granular settings. The scanner detects the individual elements by monitoring the DOM and determining whether clicking the button causes the notice element to disappear or a settings dialogue containing checkboxes to appear. 

%\subsection{Tracking and Cookie Synchronization}
%
%To detect tracking, we match each third-party request against the Disconnect tracker list~\cite{disconnect_inc_tracker_2015} and whether the request occurred before or after interaction with the detected consent notice. 
%In addition to the request, we also log the domain, the company name behind the domain, and the tracker category according to the Disconnect list.
%
%We use the method of Sanchez-Rola et al.~\cite{sanchez-rola_can_2019} to identify cookie synchronization. They used zxcvbn~\cite{wheeler_zxcvbn_2016}, a password strength estimation metric, to determine whether a cookie has enough information to identify a user uniquely. According to their work, a score of 9 is sufficient (zxcvbn provides the score as a decadic logarithm, meaning that a score of 9 means that a password/user is distinguishable between 9 billion users), so we proceeded with this value. 
%We log the request, the zxcvbn score, the cookie value, and the third-party domain to which the request synchronizes the cookie value. 
%

\section{Results}\label{sec:results}

\subsection{General Properties of the Datasets}

We scanned the top 10,000 sites on the Tranco list as of 2023-03-01 a total of three times. 
Of these scans, we randomly sampled 1,000 successful scans (we considered a scan successful if it returned a result in at least two out of three scans). 
We ran our scans between 2023-03-15 and 2023-03-17. 
As the Tables~\ref{tab:general_properties_of_the_scans} and~\ref{tab:reasons_for_scanner_failures} show, our results proved to be stable over time, yielding a similar number of failed scans, with the number of timed-out and unresolvable domains remaining constant. 

\begin{table}
\centering
\caption{\label{tab:general_properties_of_the_scans} General Properties of the Scans}
\begin{tabular}{l c c c}
\toprule
\textbf{Metric} 		& \textbf{Scan 1} & \textbf{Scan 2} & \textbf{Scan 3}\\
\midrule
Successful			& 9784				& 9777				 & 9764\\
Proportion			& 98\%				& 98\%				 & 98\%\\[0.3em]
%Failed				& 216				& 223				 & 236\\					
Notice Detected		& 3594				& 3572				 & 3572\\
Proportion			& 37\%				& 37\%				 & 37\%\\[0.3em]
Avg. Scan Duration 	& 62s				& 63s				 & 63s\\
\bottomrule
\end{tabular}
\end{table}

\begin{table}
\centering
\caption{\label{tab:reasons_for_scanner_failures} Reasons for Scan Failures}
\begin{tabular}{l c c c}
\toprule
\textbf{Metric}				& \textbf{Scan 1} 	& \textbf{Scan 2} 	& \textbf{Scan 3}\\
\midrule
\textbf{DNS Not Resolved} 	& 1503					& 1505						& 1508\\
\textbf{Not Reachable}		& 166					& 168						& 168\\
\textbf{Scanner Failure}		& 216					& 223						& 236\\
\textbf{Timeout}				& 570					& 561						& 566\\
\bottomrule
\end{tabular}
\end{table}
\subsection{Notice Detection Performance}\label{sec:banner_detection_performance}

\begin{table*}
\centering
\caption{\label{tab:sample_performance_metrics} Performance of the various detection methods (469 Notices, 531 No Notice)}
\begin{tabular}{l c c c c c c c}
\toprule
\textbf{Detection Method} 			& \textbf{TP}	& \textbf{FP}	& \textbf{FN}	& \textbf{TN} & \textbf{Precision} & \textbf{Recall} & \textbf{F1}\\
\midrule
\textbf{Method 1: DOM Tree Walking} 	& 448 & 14 & 21 & 517 & 0.97 & 0.96 & 0.96\\
\textbf{Method 2: Perceptive} 		& 434 & 14 & 35 & 517 & 0.97 & 0.93 & 0.95\\
\textbf{Method 3: EasyList Cookie} 	& 328 & 2 & 141 & 529 & 0.99 & 0.70 & 0.82\\
\textbf{Method 3: I don't care about Cookies} & 302 & 1 & 167 & 530 & 1.00 & 0.65 & 0.79\\
\textbf{Method 4: BERT Detector} & 270 & 8 & 199 & 523 & 0.97 & 0.58 & 0.73\\
\bottomrule
\end{tabular}
\end{table*}
For our analysis, we focus on precision and recall. 
In our context, \emph{precision} corresponds to the following: given the set of results obtained with a method, how many of these results are correct, i.\,e., they match the actual notices on websites according to our ground truth (for websites without a notice according to the ground truth, the result indicates that there is no notice); whereas \emph{recall} corresponds to the following: given the ground truth of all websites that do have a notice, for how many of those websites does the method find the notice correctly.

Table~\ref{tab:sample_performance_metrics} shows the results of our analysis of 1,000 randomly selected sites from the datasets after obtaining ground truth via manual inspection. 
As expected, the manually maintained filter lists ``EasyList Cookie'' and ``I don't care about cookies'' yield the highest precision. However, they recall fewer notices than the DOM Tree Walking and Perceptive method. 
The BERT detection method recalls the least number of notices and achieves a similar precision compared to the Perceptive and DOM Tree Walking detection method. This means that surprisingly, our keyword-based methods manage to achieve a similar precision compared to the BERT detection method, while managing to recall more consent notices, especially ones that are not in English. 
However, upon manual inspection and manually classifying the texts of the notices that the BERT Detector did not manage to detect, it turned out that of the 198 notices that the detector failed to recognize, 104 (52\,\%) were not in English, meaning that an English language model would have no chance of classifying them correctly. 
As for the remaining 95 (48\,\%) English notices, when feeding them to the BERT model manually, it managed to classify them correctly, which means that the heuristic of selecting the first and last three elements as well as all elements with a positive z-index needs to be improved. 

The Perceptive and DOM Tree Walking method manage to recall the most notices, with the DOM Tree Walking method outperforming the Perceptive method mainly because it performs a more generalized search. 
Due to their keyword-based nature, these approaches yield the highest number of FPs, while almost all missed notices were in a foreign language.

\paragraph{Quality of Detection}
The quality of the detection result (e.\,g., whether the detection method manages to identify the DOM node enclosing the entire notice, rather than just the headline or notice text) is another crucial aspect besides the question of whether a notice was detected or not. 
The BERT Detector delivers the best results because it is text-based and therefore detects all additional text and buttons, even if they are visually offset in the actual notice. 
The quality of the List-based Detection results is also good. 
However, in some cases, the CSS selector points to the modal parent element of the notice, which makes it impossible to collect clickable elements from it. 

The DOM Tree Walking method sometimes fails to capture all the notice elements because it stops after detecting the first button. 
In contrast, the Perceptive detection method fails to capture the buttons of a notice if they are visually offset. 
Choosing a different heuristic, which chooses the largest contour that includes the node location and is smaller than the entire website screenshot would address this issue, at least in the current dataset. However, we did not realize this during our initial exploratory scans. 
 
In summary, the Perceptive and DOM Tree Walking method manage to detect the largest number of notices, while the BERT Detector gives the most accurate results regarding notice dimensions. 
However, the Perceptive detector could be improved to match the performance of the BERT Detector, at least for the edge cases in the current dataset. In future work, we plan to apply this optimization to obtain a higher quality dataset. 

\subsection{Dark Patterns}
Our dark pattern analysis aimed to assess whether an automated scanning tool can detect obstruction (hiding a decline option one layer deep or offering no decline option at all) and interface interference (making an option more noticeable to the user by choosing a different color for the button). 

We tried to detect the presence of an option to decline on the first layer of the notice by comparing the structural similarity of the screenshots after pressing a button and before interacting with the website, as ``accept'' and ``decline'' should both lead to the disappearance of the notice, thus resulting in an identical similarity score for two screenshots. 
\begin{table}[ht]
\centering
\caption{\label{tab:opt_out_and_color_detection} Decline and Color Detection. The ``\&'' signifies the intersection of elements that are in the ground truth and were automatically detected.}
\begin{tabular}{l c c c}
\toprule
 \textbf{Metric}								& \textbf{Total}	& \textbf{Fraction}\\
\midrule	
Decline (Ground Truth)						& 246			& 52\%				\\
Decline (Ground Truth \& Automatic)			& 93				& 38\%				\\
Decline and Different Color (Ground Truth)	& 113			& 46\%				\\
Different Color (Ground Truth)				& 329 			& 70\%				\\
Different Color (Ground Truth \& Automatic)	& 249			& 76\%				\\
\bottomrule
\end{tabular}
\end{table}

Table~\ref{tab:opt_out_and_color_detection} shows the results of our analysis. 
The detection of decline options proved to be relatively poor according to the ground truth. 246 of the 469 (52\,\%) notices had an option to explicitly decline or close the notice. 
Of these, the detector managed to identify only 93 (38\,\%). 
Manual inspection revealed that the similarity scores of the screenshots differed slightly. 
In most cases, this was due to dynamic advertising, dynamic page content, or additional call-to-action pop-ups, for example, to subscribe to a newsletter. 

Out of the 246 notices with a decline option, 113 had a different color (e.g., different from the ``accept'' option or a small ``x'', usually located in the corner of the banner). 
In hindsight, it would have been better to compare the similarity of the actual web page text to a threshold to address this issue. 

In contrast to that, color detection performed much better: 
Out of the 329 (70\,\%) notices featuring different colored buttons, the scanner detected 249 (76\,\%). 
In the remaining cases, the scanner could not capture all clickable elements of the notice due to the limitation of the detection approaches mentioned in Section~\ref{sec:banner_detection_performance}. 

Of the 113 notices with different colors for accepting and declining tracking, only two presented the decline option as a link within the notice text. 
In addition, three notices presented the decline option as a link in the bottom left corner of the notice, and three displayed it as a link in the top right corner. 
Surprisingly, one notice highlighted the decline option instead of the accept option. 

In conclusion, while highlighting the accept option over the decline option is standard, presenting the decline option as a link is rare. 

\section{Discussion}\label{sec:discussion}

We assessed four consent notice detection methods on 1,000 websites randomly sampled from the top 10,000 websites listed by Tranco. 
Our results show that manually maintained filter lists (``EasyList Cookie'' and ``I don't care about Cookies'') provide the highest precision but recall fewer consent notices compared to the DOM Tree Walking and Perceptive method. 
Surprisingly, the BERT detection method has a high precision but low recall for English banners, which can be attributed to the heuristic used for selecting candidate elements rather than the model's performance. 
When presented with the missed English banner elements manually, the BERT model correctly classifies all of them.

Our keyword-based methods (DOM Tree Walking and Perceptive method) achieve similar precision to the BERT classifier while recalling more consent notices, particularly non-English ones. 
This suggests that there is a trade-off between precision and recall, where our DOM Tree Walking and Perceptive method yield the highest recall, but also deliver the highest amount of false positives. 
However, we also note that the Perceptive method could be improved by selecting the largest contour containing the word “cookie” instead of the smallest. This would improve detection when examining notices with visually separated elements. FPs could also be limited by discarding results with contours that span the screenshot of the whole website, indicating that ``cookie'' is located in the website text instead of a notice.

The detection of decline options performs relatively poor due to the dynamic nature of web content, with only 38\% of banners with a decline option being correctly identified. 
In contrast, color detection performs relatively well, detecting 76\% of notices featuring different colored buttons.

\textit{Limitations.} To ensure accurate results, we scanned our list of websites three times, each time shuffling the list of websites to reduce the chance of triggering DDoS protections of Content Delivery Networks. 
We also instrumented a real web browser with a regular user agent to ensure our scanner appeared as a normal website visitor. 
However, during our analysis, we noticed that at least one website displayed a Cloudflare error message denying access to our scanner because our IP address originated from ASN (Autonomous System Number) 680, the ASN of the German research network~\cite{reseaux_ip_europeens_network_coordination_centre_ripe_2002}. 
Although this was the only case in our evaluated scan results, we do not know whether other sites changed their behavior when interacting with our scanner. 
However, this limitation applies to other web measurement studies, especially when using virtual machines in data centers with publicly known IP ranges. 
Further research may require distributed scanning using IP addresses from residential networks to prevent them from being blocked or served other website versions, which is especially critical when examining GDPR compliance. 

Furthermore, our scanner missed a few notices since they were located in shadow DOMs, which our scanner could not access. 
We attributed those to the FPs since our scanning infrastructure could not detect them, despite some of them containing the word ``cookie''. 

\section{Conclusion}\label{sec:conclusion}

In this paper, we present and compare different methods for the detection of consent notices. 

As our results show, our keyword-based methods manage to recall the most notices, but also return the highest amount of FP. The detection method using the BERT model provides high precision, but only for English notices. The surprisingly low recall is a result of the rather arbitrary sampling method of selecting all elements with a positive z-index as well as the first three and last three elements of the body as candidate elements, rather than a failure of the trained model itself, which managed to correctly classify all missing English notice texts when presented with them manually. 

While extracting the color of the clickable elements worked well, detecting the presence of decline options by comparing the similarity of the screenshots after pressing the buttons did not. This was mostly impossible due to the dynamic nature of the web content. Comparing the similarity of the page texts using a threshold value might be more reasonable. 

We also note that while using different colored buttons to highlight the accept option is very common, disguising the decline option as a link is rare.

To our knowledge, we are the first to evaluate consent notice detection methods systematically. 
We publish the code of \emph{cookiescanner} and our dataset, which we annotated with ground truth information about the displayed consent notices. 
This allows other researchers to replicate our results and to re-use our scanning architecture for future studies on consent notices. The scanner is not limited to detecting the use of dark patterns, which we have attempted to do.  Instead, it is a useful tool for many research questions about the websites, for instance whether users are being tracked before giving consent. 
Furthermore, \emph{cookiescanner} will enable researchers to conduct studies that necessitate an automated interaction with consent notices, e.\,g., for deep scans of the behavior of websites. 
 
%\newpage
\bibliographystyle{ACM-Reference-Format}
%% Declare bibliography sources (one \addbibresource command per source)
\bibliography{bibtex.bib}
%\printbibliography
\end{document}